\documentclass[reprint,pra,superscriptaddress,notitlepage,longbibliography]{revtex4-2}

\usepackage{graphics,epsfig,amsfonts,amssymb,amsmath,color}
\usepackage[normalem]{ulem}
\usepackage{bm,bbm}
\usepackage[mathcal]{euscript}
\usepackage{color}
\usepackage{float}
\newcommand{\bra}[1]{\ensuremath{\langle#1|}}
\newcommand{\ket}[1]{\ensuremath{|{#1}\rangle}}

\begin{document}

\title{Long-range two-hybrid-qubit gates mediated by a microwave cavity with red sidebands}
\author{J. C. Abadillo-Uriel}
\affiliation{Department of Physics, University of Wisconsin-Madison, Madison, WI 53706, United States}
\affiliation{Universit\'e  Grenoble  Alpes,  CEA,  IRIG-MEM-LSIM,  38000  Grenoble,  France}
\author{Cameron King}
\affiliation{Department of Physics, University of Wisconsin-Madison, Madison, WI 53706, United States}
\affiliation{Microsoft Quantum, 1 Microsoft Way, Redmond, WA, 98052, USA}
\author{S. N. Coppersmith}
\affiliation{Department of Physics, University of Wisconsin-Madison, Madison, WI 53706, United States}
\affiliation{School  of  Physics,  University  of  New South Wales, Sydney NSW 2052, Australia}
\author{Mark Friesen}
\affiliation{Department of Physics, University of Wisconsin-Madison, Madison, WI 53706, United States}

\begin{abstract}
Implementing two-qubit gates via strong coupling between quantum-dot qubits and a superconducting microwave cavity requires achieving coupling rates that are much faster than decoherence rates. 
Typically, this involves tuning the qubit either to a sweet spot, where it is relatively insensitive to charge noise, or to a point where it is resonant with the microwave cavity. 
Unfortunately, such operating points seldom coincide.
Here, we theoretically investigate several schemes for performing gates between two quantum-dot hybrid qubits, mediated by a microwave cavity. 
The rich physics of the quantum dot hybrid qubit gives rise to two types of sweet spots, which can occur at operating points with strong charge dipole moments.
Such strong interactions provide new opportunities for off-resonant gating, thereby removing one of the main obstacles for long-distance two-qubit gates.
Our results suggest that the numerous tuning knobs of quantum dot hybrid qubits make them good candidates for strong coupling.
In particular, we show that off-resonant red-sideband-mediated two-qubit gates can exhibit fidelities $>$95\% for realistic operating parameters, and we describe improvements that could potentially yield gate fidelities $>$99\%.
\end{abstract}
\maketitle

\section{Introduction}
Semiconductor-based quantum technologies are among the most promising platforms for large-scale, universal quantum computing~\citep{Kane:1998p133, loss1998quantum}. 
Their key strengths include the possibility of leveraging advances in the semiconductor electronics industry, a highly adaptable design space that enables a wide variety of qubits and gating schemes~\citep{Petersson:2010p246804, Medford:2013p654, Shulman:2012p202, Shi:2012p140503}, attractive materials properties, and the potential to modify and improve qubit coherence~\citep{YonedaNatNano2018, yang2020operation, malinowski2017notch, medford2013resonant}. 
Over the past 20 years, many experiments have sought to exploit these strengths, and high-fidelity, single-qubit gates have been demonstrated~\citep{Kim:2015p15004,yang2019silicon,Yoneda:2018p102}. 

Recent progress has also made it possible to improve the fidelity of two-qubit gates, including short-range gates between nearest neighbors~\citep{Li:2015p7681, huang2019fidelity, Nichol:2017p1, Hendrickx_Nature2020}, and long-range gates operating over much greater distances~\citep{sigillito2019coherent, borjans2020resonant}.
Short-range gates typically exploit either capacitive coupling~\citep{Shulman:2012p202, Nichol:2017p1} or exchange interactions~\citep{loss1998quantum}, while long-range gates exploit spin shuttling~\citep{Thalineau:2012p103102, mills2019shuttling} or microwave cavities~\citep{StockklauserPRX2017,bruhat2018circuit, Mi:2018p599, Samkharadze:2018p1123}. 
In all cases, the gate fidelities are improved by increasing the inter-qubit coupling strength while minimizing the effects of environmental noise.
Here, we focus on long-range gates between qubits coupled through a microwave cavity.
Experimentally, such technology has already been used to achieve strong coupling between microwave photons and single-spin qubits~\citep{landig2019coherent, borjans2020resonant, WoerkomPRX2018, scarlino2018coherent}.
The resulting inter-qubit coupling speed is proportional to the charge dipole moment of the qubits. 
Hence, any qubits that can be manipulated quickly, via AC electric fields, are also good candidates for coupling to a microwave resonator~\citep{abadillo2019enhancing, Benito_Preprint19}. 

In this work, we focus specifically on quantum-dot hybrid qubits (QDHQs)~\citep{Shi:2012p140503, Koh:2013p19695, Kim:2014p70}.
The QDHQ is controlled fully electrically and does not require the presence of magnetic fields or magnetic field gradients that can interfere with the operation of a superconducting cavity. 
It can be tuned into a charge-qubit working regime, where it has a strong dipole moment but exhibits short coherence times~\citep{Kim:2015p243}, or a spin-valley qubit regime, where the qubit has a weaker dipole moment but is better protected from the environment~\citep{Thorgrimsson:2017p32}. 
AC gating of the QDHQ yields high-fidelity single-qubit gates~\citep{Kim:2015p15004,Thorgrimsson:2017p32}. 
Moreover, the QDHQ has many tuning knobs that can be adjusted to improve gate fidelities~\citep{Yang:2020}.
The qubit therefore shows promise for strong coupling to a microwave cavity.

To explore theoretically the coupling between a QDHQ and a microwave cavity, we map out the most interesting working regimes and estimate their two-qubit gate fidelities.
Our calculations indicate that it is most promising
to use
%We show that 
red-sideband transitions, which can be implemented between a qubit and a resonator while remaining off-resonant~\citep{BlaisPRA2007, beaudoin2012first, StrandPRB2013, SrinivasaPRB2016}. 
We propose a particular pulse sequence formed of only $Z$ rotations and red-sideband transitions, which allows us to perform effective, two-qubit, Controlled-Z (CZ) gates between distant qubits without making use of blue sidebands or intermediate leakage states, in contrast with Refs.~\citep{cirac1995quantum, childs2000universal, haack2010resonant}.
We explore the possibility of driving these gates using any one of the available tuning knobs for QDHQs: the double-dot detuning parameter, the tunnel coupling, or the valley splitting. 
As is the case for any qubit implementation, each of these methods has its own strengths and challenges. 
We describe the potential benefits of each scheme, concluding that fidelities above 95\% should be achievable for current, state-of-the-art devices.
Moreover, assuming reasonable advances in device technology, we predict that fidelities above 99$\%$ should be within reach. 

The paper is organized as follows.  
Section~II introduces a model two-level qubit, which allows us to describe the physics of longitudinal and transverse driving more pedagogically. 
We then adapt the model more specifically to the case of QDHQs. 
In Sec.~III, we describe the simulation methods used to characterize our two-qubit gates. 
In Sec.~IV, we present the results of gate simulations. 
Finally, in Sec.~V, we discuss the applicability and prospects for the proposed gating schemes.

\begin{figure*}[t]
\includegraphics[width=2\columnwidth]{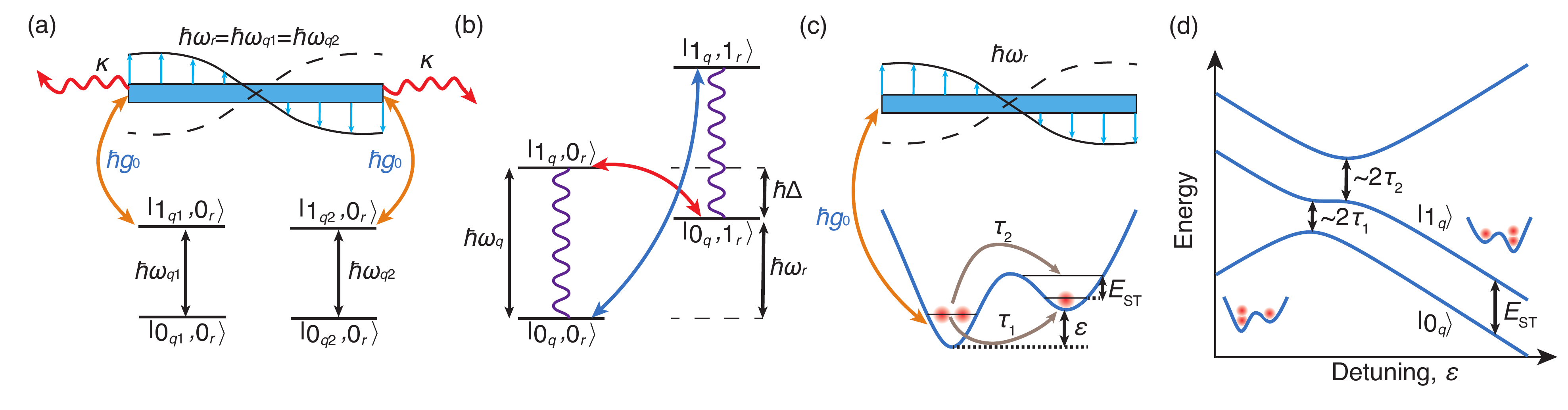}
\caption{Schematics of methods for coupling qubits via a resonator. (a) Cartoon depiction of the static qubit-cavity coupling scheme. 
For resonant coupling, two qubits with frequency $\omega_{q1}$, $\omega_{q2}$ are tuned to be in resonance with each other and with the resonator at frequency $\omega_r$. 
Interactions between the qubits and the resonator photons, $\hbar g_0(a+a^\dagger)$, with bare coupling frequency, $g_0$, are indicated by orange arrows. 
Photon loss with cavity decay rate $\kappa$ is indicated by red arrows. 
Two-qubit gates based on static coupling have relatively low fidelities: resonant gates suffer from photon leakage while dispersive gates are slow.
(b) Cartoon depiction of the sideband driving scheme for two-qubit gates. 
One of the qubits is driven, yielding Rabi oscillations (purple) of frequency $\Omega_R$.
When $\Omega_R$ equals the qubit-cavity detuning, $\Delta=\omega_q-\omega_r$, a red sideband transition occurs (red arrow), and when $\Omega_R=-\Delta$, a blue sideband transition occurs (blue arrow). 
(c) Cartoon depiction of a quantum-dot hybrid qubit (QDHQ) electrostatically coupled to a superconducting cavity. 
QDHQ model parameters include the double-dot detuning bias $\varepsilon$, the tunnel couplings $\tau_1$ and $\tau_2$, and the singlet-triplet splitting of one of the dots $E_\text{ST}$. 
(d) The lowest three energy levels of a QDHQ as a function of detuning. 
The two lowest states span the qubit subspace. 
The two anticrossings are associated with the tunnel couplings $\tau_1$ and $\tau_2$.
For large, positive detuning, the qubit frequency tends to $E_\text{ST}$. 
The insets depict the ground state charge configurations in the asymptotic detuning regimes.}
\label{fig1}
\end{figure*} 

\section{Model}
In this section we describe our theoretical model for QDHQs coupled to a superconducting cavity.
To clarify the difference between transverse and longitudinal driving, we begin by considering a generic, two-level system defined by the Hamiltonian $H_q=(\hbar\omega_q/2)\sigma_z$, where $\omega_q$ is the qubit frequency and $\sigma_z$ is a Pauli matrix.
We also describe the cavity as a quantum harmonic oscillator with the Hamiltonian $H_r=\hbar\omega_ra^\dagger a$, where $\omega_r$ is the resonator frequency and $a^\dagger$ ($a$) are microwave photon creation (annihilation) operators. 
Both the cavity and the qubits are subject to decoherence: photons leak out of the cavity at a rate $\kappa$, while the decoherence rate of the qubit depends on the specific qubit implementation, as well as the tuning and driving mechanisms.
We then adapt our model to the specific case of QDHQs, and identify the sweet-spot working regimes at which qubit decoherence rates are minimized.
Finally, we describe appropriate Hamiltonians and protocols for coupling the qubits through a resonator.
Numerical simulations of the two-qubit gates are described in later Sections of the paper.

\subsection{Two-qubit gates} \label{sec:2Qgates}
We first describe the qubit-cavity coupling schemes considered in this work. 
The simplest schemes do not involve microwave driving -- the qubit is either tuned into resonance with the cavity, or away from resonance (i.e., dispersively).
We have simulated both of these cases for QDHQs, obtaining poor gate fidelities:
resonant coupling causes excessive leakage in the form of multi-photon excitation of the resonator, while dispersive coupling is too slow, causing qubit decoherence.
After exploring a wide range of parameters, we were not able to find an acceptable operating regime for such undriven qubits. 
We therefore do not report those results here.
Instead, we focus on microwave driving. 

We consider several protocols based on microwave driving of the qubits.
These schemes may be implemented resonantly or dispersively, similar to undriven gates.
In the resonant case, the qubit and cavity states are strongly hybridized, and the main challenge for coherent operation is photons leaking out of the cavity at the rate $\kappa$.
This process, known as Purcell decay, causes the qubit to decohere at the rate $\gamma_\text{Purcell}\approx g^2\kappa/(\Delta^2+\kappa^2/4)$, where $\hbar g$ is the effective qubit-cavity coupling and $\hbar\Delta\equiv \hbar\omega_q-\hbar\omega_r$ is the qubit-cavity detuning~\citep{HouckPRL2008}.
Purcell decay can be suppressed by simply detuning the qubit away from the cavity.
Although this reduces gate speeds, we can compensate by driving the qubit.
In the context of QDHQs, we have considered several such driving protocols.
Here, we focus on the one that is found to yield the highest two-qubit gate fidelities -- a red-sideband scheme~\citep{SrinivasaPRB2016} -- for which the native gates are CZ. 

The gates considered here can be divided into two categories, transversely or longitudinally driven, depending on the form of qubit driving.
In the current subsection, we describe generic driving Hamiltonians, which can apply to many different types of qubits.
In the next subsection, we specialize to the case of QDHQs.

\subsubsection{transversely driven gates} \label{sec:transverse}
The sideband transitions depicted in Fig.~1(b) are promising tools for mediating interactions between quantum dot spin qubits~\citep{SrinivasaPRB2016}. 
Here, the qubits and cavity are intentionally detuned, to suppress the Purcell effect.
To explain the scheme, we first consider the undriven Hamiltonian,
\begin{equation}
H_\text{static}=\hbar\omega_ra^\dagger a+(\hbar\omega_{q}/2)\sigma_z+\hbar g(a+a^\dagger)\sigma_x ,
\label{eq:HDC}
\end{equation} 
where the qubit-resonator coupling, $\hbar g(a+a^\dagger)\sigma_x$, is appropriate for an electrostatically coupled quantum-dot charge qubit~\cite{SrinivasaPRB2016}.
For simplicity, we have included only one qubit in this expression, since only one qubit at a time couples to the resonator in the two-qubit protocols considered here.
However, the coupling parameter $g$ is tunable, as we discuss below; therefore, if desired, different qubits could be coupled in series to the same resonator, via the same Hamiltonian.

Next, we add a transverse driving term, which couples to the $\sigma_x$ spin component of the qubit:
\begin{multline}
H_\text{driven}=\hbar\omega_r a^\dagger a 
+\frac{\hbar\omega_{q}}{2}\sigma_z+\hbar g(a+a^\dagger) \sigma_x  \\
 +d_{01}A(t)\cos(\omega t+\phi)\sigma_{x} .
\label{eq:HAC}
\end{multline}
Here, $A(t)$ is the envelope of the drive, expressed in units of energy, $d_{01}$ is the dimensionless, transverse dipole moment, defined below, $\omega$ is the frequency of the drive, and $\phi$ is its phase.
The physical mechanism generating the drive depends on the type of qubit and will be specified for QDHQs later in this section.
If the qubit is driven resonantly ($\omega=\omega_{q}$), the corresponding Rabi frequency is given by $\Omega_{R}=d_{01}A/\hbar$.

Sideband transitions are achieved by choosing special values for the driving amplitudes, as depicted in Fig.~1(b).
To see this, we move to a frame that is co-rotating 
%with both the qubits and the resonator 
at the driving frequency $\omega=\omega_q$, as defined by the transformation~\cite{SrinivasaPRB2016}
\begin{equation}
U(t)=\exp \Big[-i\frac{\omega_qt}{2}\sigma_z-i(\omega_qt)a^\dagger a \Big] .
\end{equation}
The resulting Hamiltonian has a form similar to Eq.~(\ref{eq:HDC}), with an effective cavity frequency given by $\omega_{r,\text{eff}}=\Delta$ and an effective qubit frequency given by $\omega_{q,\text{eff}}=\Omega_{R}$.
Thus, by adjusting the Rabi frequency such that $\Omega_{R}=\pm \Delta$, we can bring the qubit into resonance with the cavity, inducing a qubit-photon transition.
Moving to a frame rotating at frequency $\Omega_{R}=\Delta$ and applying a rotating wave approximation, we obtain the effective interaction for the red ($-$) sideband transition~\citep{Blais:2004p1122}:
\begin{equation}
H_{-}\approx
\frac{\hbar g(t)}{2}\left[e^{i\phi}\sigma_{10}a+e^{-i\phi}\sigma_{01}a^\dagger\right],
\label{eq:red}
\end{equation}
where we define $\sigma_{01}=\sigma_+$ and $\sigma_{10}=\sigma_-$.
For $\Omega_{R}=-\Delta$, we obtain the effective interaction for the blue ($+$) sideband transition:
\begin{equation}
H_{+}\approx
\frac{\hbar g(t)}{2}\left[e^{-i\phi}\sigma_{10}a^\dagger+e^{i\phi}\sigma_{01}a\right].
\label{eq:blue}
\end{equation}
We see that $H_-$ contains the conventional ``rotating" terms of the cavity coupling, while $H_+$ contains the ``counter-rotating" terms.

By driving the qubit in this way for time $t$, we obtain the unitary gate operations $S_{-}(gt,\phi)=e^{-iH_{-} t/\hbar}$ or $S_{+}(gt,\phi)=e^{-iH_{+} t/\hbar}$.
The corresponding gate times are $t_g\propto 1/g$.
Compared to gate times $t_g \propto\Delta/g^2$ for conventional dispersive gates, we see that sideband gating has a clear speed advantage.
Indeed, typical sideband gate speeds are on the order of tens of ns, on par with resonant gates.
In contrast with resonant gates however, the Purcell effect is strongly suppressed for sideband gates, because they are dispersive.
Additionally, we note that, since $A(t)$ is easily tuned, the sideband resonance constraint, $\Delta=\pm \Omega_R$, is much less restrictive than the conventional resonance constraint for undriven gates ($\omega_{q}=\omega_r$).

A composite sequence based on the elemental gates $S_-$ and $S_+$ has previously been proposed for CZ gates~\citep{SrinivasaPRB2016}.
To simplify this protocol, we propose here a different gate sequence, involving just red-sideband transitions and single-qubit gates:
\begin{eqnarray}
U_\text{CZ}&=&Z^{(1)}\left(-\frac{\pi}{\sqrt{2}}\right)Z^{(2)}\left(\frac{\pi}{\sqrt{2}}\right)S_-^{(2)}(\pi,\pi)
\label{eq:seq} \\ && \hspace{-.5in} \times S_-^{(1)}(\pi/2,0)S_-^{(1)}(\pi\sqrt{2},\pi/2)S_-^{(1)}(\pi/2,\pi)S_-^{(2)}(\pi,0), \nonumber
\end{eqnarray}
where $Z^{(j)}(\theta)$ is a $Z$ rotation of angle $\theta$, acting on qubit $j$, and qubit indices are also included for the sideband gates $S_-^{(j)}(gt,\phi)$. 
Neglecting the $Z$ rotations, which are very fast, we estimate the resulting CZ gate time to be $t_\text{CZ}\approx (3+\sqrt{2})\pi/g$, which is the same as the previously studied red-and-blue-sideband protocol. 
By eliminating the blue sideband transitions however, the sequence becomes especially attractive for implementations where blue sidebands are slower than red sidebands \citep{beaudoin2012first, StrandPRB2013}.

\subsubsection{Longitudinally driven gates}
The qubit may alternatively be driven by coupling to the $\sigma_z$ component of the spin.
This is equivalent to driving the qubit energy splitting, and is referred to as longitudinal or parametric driving.
As for transverse gates, the scheme can also be used to implement red-sideband transitions between a qubit and a superconducting cavity~\citep{beaudoin2012first, StrandPRB2013}. 

To describe the coupling, we add a driving term to the qubit frequency in Eq.~(\ref{eq:HDC}): $\omega_{q}(t)=\overline\omega_{q}+(\delta\omega/2)\cos(\omega t+\phi)$, where $\overline\omega_{q}$ is the average frequency of the qubit, $\delta\omega$ is the driving amplitude, and $\omega$ is the longitudinal driving frequency.
The driving Hamiltonian then becomes
\begin{equation}
H_\text{driven}=\hbar\omega_ra^\dagger a 
+[\hbar\omega_{q}(t)/2]\sigma_z+\hbar g(a+a^\dagger)\sigma_x.
\label{eq:Hparam}
\end{equation} 
Moving to a frame defined by 
\begin{multline}
    U(t) = \exp\Big[-i\frac{\overline\omega_q t}{2}\sigma_z-i\frac{\delta\omega}{2\omega}\cos(\omega t+\phi)\sigma_z \\
    -i(\omega_rt) a^\dagger a\Big],
\end{multline}
we find that the effective interaction for a longitudinally driven red-sideband transition is given by
\begin{eqnarray}
&& \hspace{-0.1in} H_-\approx 
\label{eq:Hpred} \\ &&
\hbar g \sum_{n=0}^\infty \left[(-i)^n J_n \left(\frac{\delta\omega}{\omega}\right)e^{i(n\omega-\Delta) t+i\phi}a^\dagger \sigma_{01} + \text{h.c.} \right],
\nonumber
\end{eqnarray} 
where we have used the Anger-Jacobi expansion~\cite{Cuyt:2008}.
Here, the approximate equality indicates that we have dropped the counter-rotating (blue-sideband) terms in the expression; we have verified that this approximation is accurate in the operating regimes of interest for this work.
As before, the interaction between the qubit and the resonator is dispersive; however the coupling can now be initiated by applying a driving frequency $n\omega=\Delta$ (a red sideband constraint), which yields a renormalized coupling of strength $\hbar g J_n(\delta\omega/\omega )$. 
The optimal (i.e., strongest) interaction is obtained when $\omega\approx \delta\omega/1.84$, corresponding to the $n=1$ term of the Anger-Jacobi expansion.
We do not consider the $n=0$ term here because it gives a significant blue-sideband contribution that complicates the qubit dynamics. 
A CZ gate is then obtained using the sequence given in Eq.~(\ref{eq:seq}).

\subsection{Quantum Dot Hybrid Qubits}
The effective qubit-resonator coupling $g$ determines the two-qubit gate speed.
It is the product of the bare coupling, $g_0$, which depends on the device geometry, and the charge-dipole matrix element, which depends on the choice of qubit and the tuning of that device.
The qubit dephasing rate also depends on these same parameters.
To weigh these two competing effects, we must consider a specific qubit model.
Here, we describe a minimal Hamiltonian for the QDHQ and use it to compute $g$. 
In the following section, we use the same Hamiltonian to characterize the QDHQ dephasing.

The QDHQ is formed in a double-quantum dot with three electrons~\citep{Shi:2012p140503, Koh:2013p19695}. 
Denoting $(n,m)$ as the charge configuration with $n$ electrons in the left dot and $m$ electrons in the right dot, we define $\varepsilon$ as the energy difference or ``detuning" between the (1,2) and (2,1) configurations, with $\varepsilon =0$ corresponding to the degeneracy point.
Note here that the double-dot detuning ($\varepsilon$) is different than the qubit-cavity detuning ($\Delta$).
In addition to charge, the system also has a spin degree of freedom, and the lowest excitation energy for a given charge configuration is approximately equal to the singlet-triplet splitting, $E_\text{ST}$, of the dot containing two electrons.
Here we adopt the convention that $\varepsilon <0$ represents the tuning regime for which the ground-state splitting is largest.
We will assume that this regime is used for readout while the $\varepsilon\geq 0$ regime is used to perform qubit gate operations. 
Since the excited spin state does not play a role in readout, and since the excitation energy is relatively large in this case, we may simply ignore the excited (2,1) spin state in a minimal model.
The resulting model is then three-dimensional (3D), with one (2,1) state and two (1,2) states.
Defining the spin basis states as $\{\ket{S,\downarrow},\ket{\downarrow,S},\sqrt{1/3}\ket{\downarrow,T_0}+\sqrt{2/3}\ket{\downarrow,T_-}\}$, the 3D Fermi-Hubbard Hamiltonian for the QDHQ becomes~\citep{Shi:2012p140503, Koh:2013p19695}
\begin{equation}
H=\begin{pmatrix}
\varepsilon/2 & \tau_1 & \tau_2 \\
\tau_1 & -\varepsilon/2 & 0 \\
\tau_2 & 0 & -\varepsilon/2+E_\text{ST}
\end{pmatrix}, \label{eq:Hhq}
\end{equation} 
where $\ket{\uparrow}$ and $\ket{\downarrow}$ indicate the spin states of the singly occupied dot, $\ket{S}$, $\ket{T_0}$, and $\ket{T_-}$ refer to spin singlet, and spin-zero or spin-polarized triplet states in the doubly occupied dot, $\tau_1$ denotes the tunnel coupling between states $\ket{S,\downarrow}$ and $\ket{\downarrow,S}$, and $\tau_2$ denotes the tunnel coupling between states $\ket{S,\downarrow}$ and $\sqrt{1/3}\ket{\downarrow,T_0}+\sqrt{2/3}\ket{\downarrow,T_-}$, as illustrated in Fig.~1(c).

The eigenvalues of Hamiltonian~(\ref{eq:Hhq}) are plotted in Fig.~1(d) as a function of the detuning parameter $\varepsilon$ for typical QDHQ settings.
The lowest two levels represent the qubit states $\ket{0}$ and $\ket{1}$, while the third level
%, which is used to mediate single-qubit gate operations, 
is a leakage state, $\ket{L}$.
Near zero detuning, the qubit states are charge-like, with a large dipole moment that can couple to the microwave cavity, but which also interacts strongly with the environment. 
Away from zero detuning, the qubit states become more spin-like, with nearly identical charge configurations that are well protected from environmental charge noise~\citep{Thorgrimsson:2017p32, Abadillo-Uriel:2018p165438}, but the small dipole moment does not couple strongly to the microwave cavity.

The electrostatic coupling between the qubit and the cavity is typically mediated by a metal top gate, fabricated above one of the dots, which taps into one end of the microwave cavity~\cite{StockklauserPRX2017}.
This coupling is proportional to the capacitive coupling $g_0$, between the qubit and the top gate, and the (dimensionless) charge dipole moment, $d_{01}$, defined below.
For sideband transitions, the qubit is driven, yielding a time-dependent dipole moment, $d_{01}(t)$.
The effective coupling between the qubit and the resonator is then given by $g(t)\approx g_0\,d_{01}(t)$.
We now calculate the dipole matrix element for a QDHQ, following Refs.~\citep{PhysRevA.95.062321} and \cite{Frees:2019p73}.

We first diagonalize Eq.~(\ref{eq:Hhq}) at a desired working point.
The diagonal Hamiltonian $H'$ is related to the original Hamiltonian $H$ by
\begin{equation}
H'=U^\dagger H U = \sum_n E_n\ket{n}\bra{n}, \label{eq:Hdiag}
\end{equation}
where $E_n$ is the energy of eigenstate $\ket{n}\in \{\ket{0},\ket{1},\ket{L}\}$. 
The unitary operator $U$ can be expressed analytically in certain limits~\citep{PhysRevA.95.062321}, but is more generally obtained by numerical diagonalization.
We refer to the diagonal basis as the qubit frame, while the original basis, corresponding to Eq.~(\ref{eq:Hhq}), defines the charge frame.
The dipole operator is naturally expressed in the charge frame as 
\begin{equation}
\hat d=\begin{pmatrix}
1/2 & 0 & 0 \\
0 & -1/2 & 0 \\
0 & 0 & -1/2
\end{pmatrix} .
\end{equation}

To compute the qubit-cavity coupling, we need to express the dipole operator in the qubit frame: $d'=U^\dagger \hat dU$.
Defining the dipole matrix elements as $d_{nm}'\equiv \bra{n}d'\ket{m}$ and the generalized Pauli matrices as $\sigma_{nm}\equiv \ket{n}\bra{m}$,
the qubit-cavity coupling becomes
\begin{equation} g_0\sum_{n,m}^{0,1,L}d_{nm}'(a+a^\dagger)\sigma_{nm} .
\label{eq:geff}
\end{equation}
Note that while the qubit-cavity coupling is purely transverse in the charge frame, $g_0d_{01}(a+a^\dagger)\sigma_x$, in the qubit frame it can have both longitudinal ($d_{00}', d_{11}'$) and transverse ($d_{01}',d_{10}'$) components.

We may investigate the limiting behaviors of the qubit-cavity coupling.
When $\varepsilon\approx 0$, the eigenstates are completely delocalized, with $\ket{0}\approx(\ket{S,\downarrow}-\ket{\downarrow,S})/\sqrt{2}$ and $\ket{1}\approx(\ket{S,\downarrow}+\ket{\downarrow,S})/\sqrt{2}$.
Here, small deviations of $\varepsilon$ from zero cause large shifts in the charge state.
We then find that $d_{01}'\approx 1/2$ is maximized, while $d_{00}'\approx d_{11}'\approx 0$, as consistent with a dominant transverse coupling.
When $\varepsilon>0$, additional longitudinal components are present in the coupling.
In the far-detuned limit, $\varepsilon\gg \tau_1,\tau_2,E_\text{ST}$, the qubits are strongly localized in the same (1,2) charge configuration, and variations of $\varepsilon$ have almost no effect on the charge state.
Consequently, the dipole matrix elements are very small, although both transverse and longitudinal components are present.
This ability to tune the dipole matrix element provides an important resource for controlling two-qubit gate operations. 
Alternatively, $g$ can be set to zero by simultaneously suppressing both the tunnel couplings, which prevents charge motion between dots.

The QDHQ qubit can also be manipulated by externally modulating the Hamiltonian parameters as a function of time.
For the case of detuning modulations, $\delta\varepsilon(t)$, the perturbation to the Hamiltonian~(\ref{eq:Hhq}) in the charge frame is given by
\begin{equation}
H_\varepsilon=\delta\varepsilon(t)\begin{pmatrix}
1/2 & 0 & 0 \\
0 & -1/2 & 0 \\
0 & 0 & -1/2
\end{pmatrix} , \label{eq:epsdrive}
\end{equation}
which we readily identify as $H_\varepsilon=\delta\varepsilon(t)\hat d$.
Transforming to the qubit frame, we then have 
\begin{equation}
H'_\varepsilon=U^\dagger H_\varepsilon U = \sum_{n,m}^{0,1,L}\delta\varepsilon(t) h_{nm}^{(\varepsilon)}\sigma_{nm},
\label{eq:depsilon}
\end{equation}
where the matrix elements for detuning driving are given by $h_{nm}^{(\varepsilon)} = d_{nm}'$.

Double-dot qubits can also be driven by modulating the tunnel couplings.
For the QDHQ, the driving Hamiltonian in the charge frame is given by
\begin{equation}
H_\tau=\delta\tau(t)\begin{pmatrix}
0 & \delta\tau_1/\delta\tau & d\tau_2/\delta\tau \\
\delta\tau_1/\delta\tau & 0 & 0 \\
d\tau_2/\delta\tau & 0 & 0
\end{pmatrix} .
\label{eq:deltatau}
\end{equation}
Here, we assume the tunnel couplings are driven simultaneously, in a fixed ratio $\delta\tau_1/\delta\tau_2$, by changing the voltage on a single top gate~\cite{Yang:2020,Borjans:2021preprint}.
Transforming to the qubit frame, we obtain the driving Hamiltonian
\begin{equation}
H'_\tau=U^\dagger H_\tau U =
\sum_{n,m}^{0,1,L} \delta\tau(t)
h_{n,m}^{(\tau)}\sigma_{nm} , \label{eq:taudrive}
\end{equation}
which defines the matrix element $h_{nm}^{(\tau)}$ for tunnel-coupling driving.

Finally, we can drive the $E_\text{ST}$ term in Eq.~(\ref{eq:Hhq}).
In this case, in the charge frame, we simply have $H_{E_\text{ST}}=\delta E_\text{ST}(t)\,\text{diag}\{0,0,1\}$, which in the qubit frame yields
\begin{equation}
H'_{E_\text{ST}}=U^\dagger H_{E_\text{ST}} U =
\sum_{n,m}^{0,1,L} \delta E_\text{ST}(t)
h_{n,m}^{(E_\text{ST})}\sigma_{nm}~,  \label{eq:ESTdrive}
\end{equation}
which defines the matrix element $h_{n,m}^{(E_\text{ST})}$.
In silicon quantum dots, $\delta E_\text{ST}$ can be modulated through the orbital confinement potential \citep{ercan2021strong} or the valley splitting, as described below. 

\subsection{Optimal Working Points} \label{sec:optimal}
Charge noise that causes the detuning parameter $\varepsilon$ to fluctuate~\citep{Dial:2013p146804} is a problem for all quantum dot qubits.
For Si quantum dots with two electrons, in most cases, it constitutes the main dephasing mechanism~\citep{Gamble:2012p035302}.
Previous work has shown that the optimal working points for single-qubit gates in a QDHQ should occur at sweet spots, where the derivative of the qubit frequency with respect to detuning vanishes~\citep{Thorgrimsson:2017p32}.
We have also shown previously that the optimal fidelities of two-qubit gates of double-dot singlet-triplet qubits mediated by a microwave cavity occur near sweet spots~\citep{abadillo2019enhancing}.
We now map out the sweet spots for the QDHQ.

Numerical computation of the parameter values of the sweet spots at which
$d\omega_q/d\varepsilon=0$, using Eq.~(\ref{eq:Hhq}), 
 reveals regimes with zero, one, or two sweet spots, depending on the values of the tunnel couplings, as indicated in Fig.~2(a).
Some examples of sweet spots are shown in Fig.~2(b).
Here, the sweet spot locations are indicated on the detuning axis, and their corresponding tunnel couplings values indicated by purple star, triangle, and square markers in Fig.~2(a).
For cases with two sweet spots (e.g., the top or bottom panels), the sweet spot nearest to zero detuning occurs in a regime where the qubit is charge-like; we therefore refer to this as a charge sweet spot (CSS).
The other sweet spot is less charge-like, and we refer to it as a %spin or 
valley sweet spot (VSS).

\begin{figure*}[t]
\includegraphics[width=2\columnwidth]{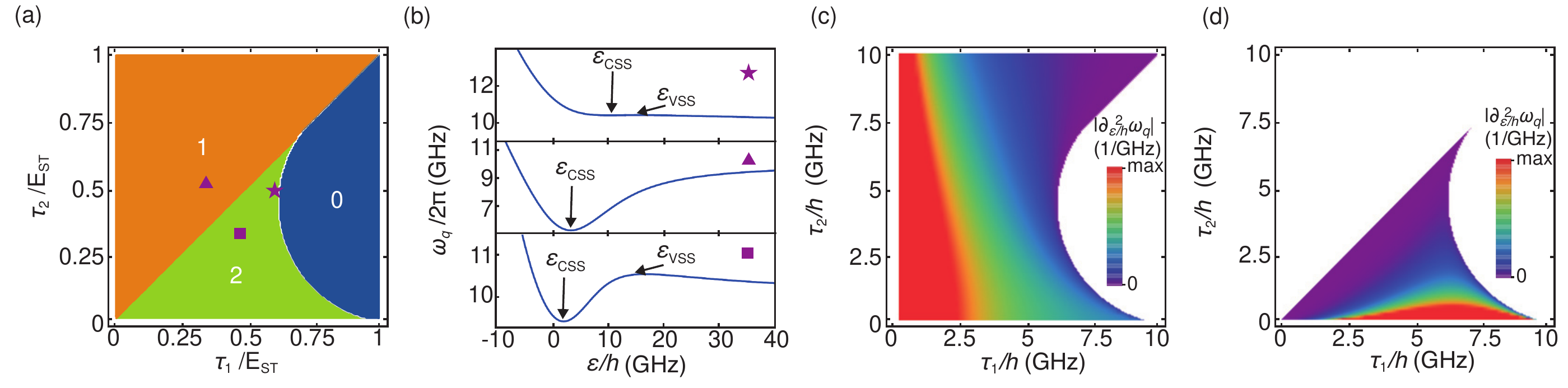}
\caption{
QDHQ sweet spots and energy dispersions.
Qubit decoherence arising from detuning fluctuations is suppressed at sweet spots, where the first derivative of the qubit frequency with respect to detuning is zero.  Having a small
second derivative of the qubit frequency with respect to detuning is also favorable for reducing the decoherence arising from charge noise.
(a) The number of available sweet spots as a function of the two tunnel-coupling parameters. 
(b) Three examples of QDHQ energy dispersions with either one (middle panel) or two (top and bottom panels) sweet spots.
The corresponding tunnel-coupling values are indicated by the star, triangle, and square markers in (a).
The star shows the location of the highest gate fidelity, as computed in Sec.~IV, corresponding to $(\varepsilon,\tau_1,\tau_2,E_\text{ST})/h=(15.59,6.18,4.9,10)$~GHz.
Each sweet spot is labeled as a charge or valley sweet spot (CSS or VSS), as described in the main text. 
(c) The second derivative of the qubit frequency with respect to detuning, evaluated at the CSS. 
(d) The same quantity, evaluated at the VSS. 
We set $E_\text{ST}/h=10$~GHz for all results shown here.}
\label{fig2} 
\end{figure*}

The CSS and VSS can both be used as working points, although they offer different resources for gate operations.
The CSS has a stronger charge dipole, since it occurs near zero detuning, which allows for faster cavity-mediated gates.
On the other hand, the qubit dispersion is typically flatter near a VSS, and its sweet spot is wider, yielding better coherence properties.
We can characterize the qubit coherence by computing the second derivative of its dispersion relation. 
These results are plotted for the CSS in Fig.~2(c), and for the VSS in Fig.~2(d).
Note that the VSS occurs only in the region with two sweet spots, while the CSS occurs in the regions with one or two sweet spots.
Since it is important for the single-qubit coherence time to be long, we only consider sweet spots as working points; we therefore exclude the curved white region on the right-hand side of Figs.~2(c) and 2(d), which contains no sweet spots.

In Figs.~2(c) and 2(d), we see that the best coherence (i.e., the lowest $\partial^2\omega_q/\partial \varepsilon^2$) is found near the boundaries between regions with different numbers of sweet spots.
At the first boundary, between the regions with one or two sweet spots (along the line $\tau_1=\tau_2$), the VSS moves out to $\varepsilon\rightarrow\infty$, where the qubit dispersion is very flat; the other sweet spot (a CSS) occurs at finite $\varepsilon$.
At the boundary between regions with zero or one sweet spot (also along the line $\tau_1=\tau_2$), there is only one sweet spot (a CSS), which moves out to $\varepsilon\rightarrow\infty$.
At the third boundary, between the regions with zero or two sweet spots, the CSS and VSS merge into a line of second-order sweet spots, where $d\omega_q/d\varepsilon = d^2\omega_q/d\varepsilon^2 = 0$.
In general, we expect this regime to be optimal for gate operations because the dispersion is extremely flat, and the dipole moment is not vanishingly small.

Finally, we note that the gate-fidelity calculations, described in later sections, include additional physics that is not reflected in Figs.~2(c) and 2(d) -- in particular, they include leakage.
As a result, it will be shown that the best fidelities are not achieved precisely at the second-order sweet spots, because the leakage effects are also most prominent along this line.
However, the dispersion is still quite flat in the vicinity of the second-order sweet spots.
Consequently, the highest fidelity working point in this work is found to occur near the triple-point, $(\tau_1,\tau_2)/E_\text{ST}=(1,1)/\sqrt{2}$~\citep{PhysRevA.95.062321}, where the regions with zero, one, and two sweet spots all converge.
This optimal working point is indicated by a purple star in Figs.~2(a) and 2(b).

\subsection{QDHQ Two-Qubit Hamiltonians} \label{sec:2QH}
The two-qubit Hamiltonians of Sec.~II.A were developed for simple two-level systems; however, they can be extended to describe the three-level QDHQ system.
First, we combine Eq.~(\ref{eq:Hdiag}) and terms involving the cavity to obtain the static Hamiltonian,
\begin{multline}
{H'}_\text{static}=\hbar\omega_ra^\dagger a
+\sum_{n}^{0,1,L}\sum_{j}^{1,2}E_n^{(j)}\sigma_{nn}^{(j)} \\
+\sum_{n,m}^{0,1,L}\sum_{j}^{1,2} g_0^{(j)}{d'}_{nm}^{(j)}(a+a^\dagger)\sigma_{nm}^{(j)},
\label{eq:resonant}
\end{multline} 
where the index $j$ labels the two qubits.
As described in Sec.~\ref{sec:2Qgates}, we then consider perturbations to the QDHQ control parameters, $\delta \varepsilon$, $\delta\tau$, or $\delta E_\text{ST}$, in the form of sinusoidal driving terms.
Denoting the time-varying envelopes of the driving terms as $A_\varepsilon(t)$, $A_\tau(t)$, and $A_{E_\text{ST}}(t)$, the full Hamiltonian in the qubit frame becomes
\begin{multline}
{H'}_\text{driven}=\hbar\omega_ra^\dagger a
+\sum_{n}^{0,1,L}\sum_{j}^{1,2}E_n^{(j)}\sigma_{nn}^{(j)} \\
\hspace{-.6in}+\sum_{n,m}^{0,1,L}\sum_{j}^{1,2}\bigg[ g_0^{(j)}\,{d'}_{nm}^{(j)}(a+a^\dagger) \\ 
+\bigg(A^{(j)}_\varepsilon\!(t)\, h^{(\varepsilon,j)}_{nm}
+A^{(j)}_\tau\!(t)\,h_{nm}^{(\tau,j)} \\
+A^{(j)}_{E_\text{ST}}\!(t)\,h_{nm}^{(E_\text{ST},j)} \bigg) \cos(\omega^{(j)}t+\phi_j) 
\bigg] \sigma_{nm}^{(j)} . \label{eq:Hdriv}
\end{multline}
For definiteness, we will always choose  $A_{\tau,\text{max}}=\sqrt{\delta\tau_1^2+\delta\tau_2^2}=\sqrt{\tau_1^2+\tau_2^2}/5$ in the simulations described in Secs.~\ref{sec:CNsims} and \ref{sec:sims}. 

Again, we can identify appropriate sideband transitions in Eq.~(\ref{eq:Hdriv}).
The QDHQ tuning parameters $\varepsilon$ and $\tau$ modify the qubit dispersion most strongly in the vicinity of $\varepsilon\approx 0$, where the matrix elements $h_{nm}^{(\varepsilon,j)}$ and $h_{nm}^{(\tau,j)}$ are predominantly transverse.
We therefore use the $\varepsilon$ and $\tau$ terms to provide transverse driving. 
These parameters can be driven independently, or in tandem, to obtain larger Rabi frequencies.
The qubit should be driven resonantly, with $\omega^{(j)}=\omega_q^{(j)}$; the modified transverse Rabi sideband constraint is then given by $h^{(\varepsilon,j)}_{01}A_{\varepsilon,\text{max}}^{(j)}+h_{01}^{(\tau,j)}A_{\tau,\text{max}}^{(j)}=\pm\Delta^{(j)}$, for red or blue sidebands.
Note that the Rabi constraint specifically picks out the transverse components of $h_{nm}^{(\varepsilon,j)}$ and $h_{nm}^{(\tau,j)}$, although other matrix components are also present in Eq.~(\ref{eq:Hdriv}).

The $E_\text{ST}$ parameter describes the singlet-triplet splitting of the dot with two electrons. 
Modulating $E_\text{ST}$ requires varying either the dot confinement potential or its valley splitting.
The former occurs naturally, any time there is a relative change in neighboring top-gate voltages, while the latter is accomplished by shifting the lateral position of the dot or varying the electric field perpendicular to the quantum well~\cite{Friesen:2007p115318,Shi:2011p233108,Abadillo-Uriel:2018p165438,PhysRevApplied.13.034068}.
In this work, unless noted otherwise, we will always assume $E_\text{ST}=10$~GHz as a typical value.
By further lowering the value of $E_\text{ST}$, we could potentially loosen the constraints on the cavity resonance frequency, as described below.
However, lower $E_\text{ST}$ also leads to increased leakage, as discussed in Appendix~A.
Higher $E_\text{ST}$ values suggest optimal cavity frequencies that are difficult to achieve in the laboratory, as also described below.
The value $E_\text{ST}=10$~GHz is therefore a convenient, intermediate, and realistic choice.

Simulations of the valley splitting in Si quantum dots suggest the maximum amplitude that can be achieved for driving of the  $E_\text{ST}$ parameter, without altering the electronic charge occupations, is approximately given by $A_\text{ST,max}\approx 4$~GHz, when $E_\text{ST}=10$~GHz (see Appendix~B).
An important benefit of using $E_\text{ST}$ as a driving parameter is that gating can still be accomplished when the qubit is tuned to the asymptotic regime, $\varepsilon\rightarrow\infty$, where the qubit dispersion is very flat.
In this regime, the matrix element components $h_{nm}^{(E_\text{ST},j)}$ are primarily longitudinal, and we therefore employ $E_\text{ST}$ for longitudinal driving.
As described in Sec.~\ref{sec:2Qgates}, the resonant condition for longitudinal driving is then given by $n\omega^{(j)}=\Delta^{(j)}$.
We note that, while non-longitudinal components of $h_{nm}^{(E_\text{ST},j)}$ are present in Eq.~(\ref{eq:Hdriv}), these yield negligible off-resonant corrections to the dynamics.

\begin{figure*}[t]
\includegraphics[width=1\textwidth]{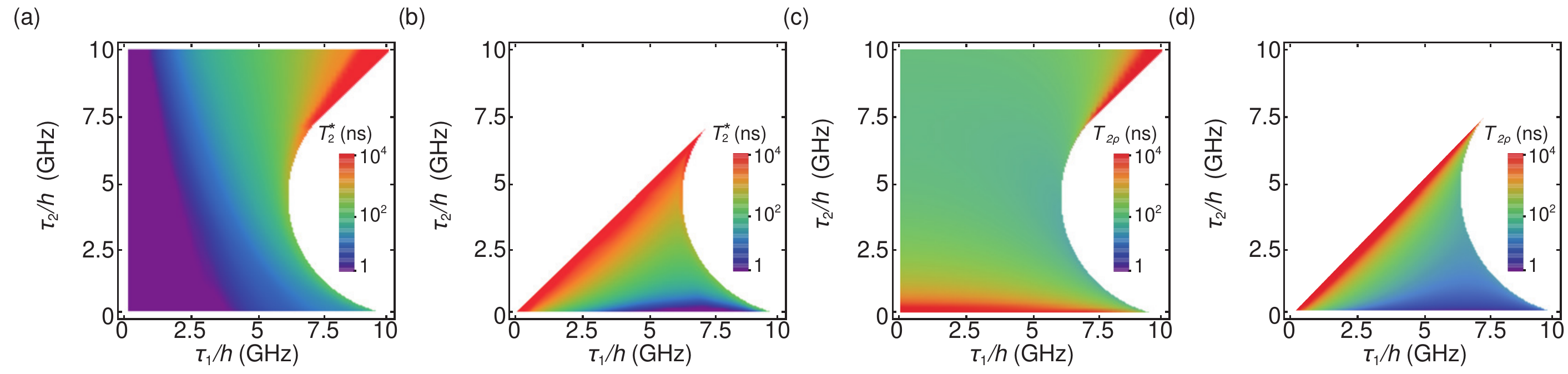}
\caption{
Computed single-qubit decoherence times due to $1/f$ charge noise, as defined in Eq.~(\ref{eq:noise}).
(a) $T_2^*$ computed at a CSS. 
(b) $T_2^*$ computed at a VSS.
(c) $T_{2\rho}$ computed at a CSS.
(d) $T_{2\rho}$ computed at a VSS.
In (c) and (d), we show the results for $\varepsilon$-driving 
with amplitude $A_\varepsilon =1$ GHz; however solutions for different values of $A_\varepsilon$ are also included in our two-qubit simulations.}
\label{newfig3}
\end{figure*}

\section{Simulation Methods}
Optimizing two-qubit gate fidelities requires balancing gate speeds and decoherence rates.
In this section, we describe the methods used to include noise in our numerical simulations.
The simulations proceed in two steps. 
First, we estimate the relevant coherence times of the system by injecting $1/f$ charge noise into the detuning parameter. 
We then incorporate these coherence times into master equations that describe the dynamical evolution of the various gate protocols.

\subsection{Charge-Noise Simulations} \label{sec:CNsims}
We consider fluctuations of the qubit detuning parameter, defined as $\varepsilon(t)=\overline\varepsilon+\delta\varepsilon(t)$, where the average detuning value, $\overline\varepsilon$, is always set to a sweet spot, as described in previous sections.
Fluctuating time series for $\delta\varepsilon(t)$ are generated using the method described in Refs.~\citep{abadillo2019enhancing} and \citep{Yang:2019}. 
Here, we adopt a $1/f$ charge-noise power spectrum, commonly observed in semiconductor qubits~\cite{paladino20141,Yoneda:2018p102}, defined as
\begin{equation}
\label{eq:noise} 
{S}_\varepsilon (\omega) = \left\{
  \begin{array}{cl}
    \frac{2\pi c_{\varepsilon}^2}{|\omega|} & \text{for}\,\, \omega_l \leq |\omega | \leq \omega_h\\
    0 & \text{otherwise} 
  \end{array}
\right. ,
\end{equation} 
where the low- and high-frequency cutoffs are given by $\omega_l/2\pi=100$~kHz, and $\omega_h/2\pi =40$~GHz. 
Unless otherwise noted, we will always employ a noise amplitude of $c_\varepsilon=0.56$~$\mu$eV, which corresponds to a noise standard deviation of $\sigma_\varepsilon=2$~$\mu$eV, through the relation $\sigma_\varepsilon = c_\varepsilon \big[2\ln(\sqrt{2\pi}c_\varepsilon/\hbar\omega_l)\big]^{1/2}$~\citep{jock2018silicon, abadillo2019enhancing}. 

To determine $T_2^*$, we simulate the QDHQ free-induction decay by considering the initial state $\ket{\psi (0)}=(\ket{0}+\ket{1})/\sqrt{2}$, which we evolve according to Hamiltonian~(\ref{eq:Hhq}), including the detuning fluctuations.
The resulting evolution is expressed in terms of the density matrix $\rho(t)=\ket{\psi(t)}\bra{\psi(t)}$.
We repeat this procedure 10$^4$ times, with different random time series, and compute the average density matrix, $\overline \rho(t)$.
$T_2^*$ is determined by fitting the off-diagonal term of this average to the form
\begin{equation}
|\overline \rho_{01}(t)|=\exp[-(t/T_2^*)^\beta]/2.
\end{equation}
Here, $\beta$ is also a fitting parameter, which is found to be close to unity over most of the parameter range. 
Since $\overline\varepsilon$ is tuned to a detuning sweet spot, the resulting $T_2^*$ is found to be a function of the tunnel-coupling parameters that determine the location of the sweet spot.

The results of these simulations are shown in Fig.~3(a) for the case where $\overline\varepsilon$ is set to a CSS, and in Fig.~3(b) where $\overline\varepsilon$ is set to a VSS.
In the first case, we obtain relatively poor coherence on the left-hand side of the plot, corresponding to small $\tau_1$ and large $\tau_2$, because the CSS is very narrow in this regime.
Better coherence is achieved in the upper-right-hand corner, near the line $\tau_1=\tau_2$, because the sweet spot occurs in the far-detuned regime, where the energy dispersion is very flat.
In the second case [Fig.~3(b)], we obtain relatively poor coherence on the bottom of the plot (small $\tau_2$, large $\tau_1$), because the VSS is very narrow.
Better coherence is again achieved near the line $\tau_1=\tau_2$, because the sweet spot occurs in the far-detuned regime.
We note that $T_2^*$ values in the range of 5-10~$\mu$s are often unphysical, since other decoherence mechanisms take over~\cite{Gamble:2012p035302}.
For both CSS and VSS, fairly good coherence is also achieved near the boundary between regions with zero and two sweet spots [see Fig.~2(a)], because this corresponds to a line of second-order sweet spots.

\subsection{Rabi-Frequency Fluctuations} \label{sec:Rabiflucts}
Charge noise can affect qubits in new ways when they are driven~\citep{Yan:2013p2337,Jing:2014p022118}.
In Ref.~\citep{abadillo2019enhancing}, we showed that fluctuations of the Rabi frequency, $\Omega_R(t)=\overline \Omega_R+\delta\Omega_R(t)$, become important when single-triplet qubits are driven.
Here we show that similar fluctuations also affect QDHQs.
The origin of the effect can be understood from Eq.~(\ref{eq:Hdriv}).
We focus on the case of $\varepsilon$-driving, because it is more challenging to apply strong $\tau$-driving or $E_\text{ST}$-driving in the laboratory, as discussed below.
After moving to the rotating frame, the Rabi frequency, $\Omega_R=A_\varepsilon(t)h_{01}^{(\varepsilon)}$, plays the role of a quantizing field.
However, as noted, the Rabi frequency is itself susceptible to charge-noise fluctuations, causing dephasing to occur in the rotating frame, over the characteristic time scale $T_{2\rho}$.
This noise mechanism is primarily quasistatic or low-frequency~\citep{Yan:2013p2337}; for double-dot qubits, it couples to the drive through fluctuations of the dipole matrix element, $d_{01}^{(\varepsilon)}(t)=\overline d_{01}^{(\varepsilon)}+ \delta d_{01}^{(\varepsilon)}(t)$.
Following Ref.~\cite{abadillo2019enhancing}, we can write the resulting contribution to dephasing in the rotating frame as
\begin{equation}
\frac{1}{T_{2\rho}}=\frac{A_\varepsilon}{4\hbar}\frac{\partial d_{01}^{(\varepsilon)}}{\partial\varepsilon}\sigma_\varepsilon.  
\label{eq:T2rho}
\end{equation}

Using the simulation method described in the previous subsection, we obtain estimates for $T_{2\rho}$, as shown in Fig.~3(c) for the case when $\overline\varepsilon$ is tuned to a CSS, and in Fig.~3(d) for the case when $\overline\varepsilon$ is tuned to a VSS.
We see that $T_{2\rho}$ is enhanced along the line $\tau_1=\tau_2$, which corresponds to the far-detuned regime, because here the dipole moment $d_{01}$ is very small, so its derivative in Eq.~(\ref{eq:T2rho}) is also very small.
In Fig.~3(c), $T_{2\rho}$ is also enhanced along the line $\tau_2=0$, which corresponds to a conventional charge sweet spot, with $\partial d_{01}^{(\varepsilon)}/\partial\varepsilon=0$.
In Fig.~3(d), the same line corresponds to a sweet spot in the transverse components of the interaction.
In the two-qubit simulations described below, we include $T_{2\rho}$ fluctuations using a master equation approach.
We note that, since the master equation is expressed in the laboratory frame, the Lindblad operator for Rabi-frequency fluctuations takes the form of a transverse decay term.

\subsection{Two-Qubit Gate Simulations} \label{sec:sims}
We now compute the gate fidelities for two-qubit CZ gates, performed by driving one of the QDHQ tuning parameters, $\varepsilon$, $\tau_1$ (and $\tau_2$), or $E_\text{ST}$. 
To simulate a given gating protocol, we adopt the master equation
\begin{multline}
\dot{\rho}=-\frac{i}{\hbar}[H_\text{driven},\rho]
+\sum_j^{1,2}\Big[ \frac{1}{2T_2^*}(\sigma_{z}^{(j)}\rho\sigma_{z}^{(j)}-\rho)  \\
+ \frac{1}{2T_1}(2\sigma_{01}^{(j)}\rho\sigma_{10}^{(j)}-\sigma_{10}^{(j)}\sigma_{01}^{(j)}\rho-\rho\sigma_{10}^{(j)}\sigma_{01}^{(j)})  \\
+ \frac{1}{2T_{2\rho}}(\sigma_{01}^{(j)}\rho\sigma_{10}^{(j)}+\sigma_{10}^{(j)}\rho\sigma_{01}^{(j)}-\rho)  \\
+ \frac{\kappa}{2}(2a\rho a^\dagger-a^\dagger a\rho-\rho a^\dagger a) \Big] , \label{eq:Master}
\end{multline}
where $\sigma_z\equiv \sigma_{00}-\sigma_{11}$, and we incorporate the coherence times obtained in the previous section.
Note that we have included a phenomenological relaxation term with $T_1=10$~$\mu$s, to account for the effects of phonons~\citep{PhysRevLett.108.046808, PhysRevB.100.035310}; this term is important when charge-noise effects are suppressed, for example in cases where the energy dispersion is very flat~\cite{Gamble:2012p035302}.
We have also accounted for photons leaking out of the microwave cavity at rates in the range of $\kappa/2\pi=0$-10~MHz~\citep{BlaisPRA2007}, and we have introduced $T_{2\rho}$ noise using the Lindblad operators described in Ref.~\citep{abadillo2019enhancing}.
The operators $H_\text{driven}$ and $\rho$ are both $9N_r$-dimensional, corresponding to $3^2=9$ two-qubit states, and $N_r=5$ photon-number modes in the resonator. 
(We have checked that cavity modes with $N_r>5$ play no role, for the qubit parameters and driving parameters considered here.)
Clearly, Eq.~(\ref{eq:Master}) does not describe all possible decoherence channels in this high-dimensional system; however, it incorporates the most important effects for two-qubit gates, including $T_2^*$, $T_1$, $T_{2\rho}$, and $\kappa$ processes, and coherent leakage to the quantum dot $\ket{L}$ states and cavity states with $N_r>1$.

We calculate the fidelities of two-qubit gates based on the $U_\text{CZ}$ pulse sequence given in Eq.~(\ref{eq:seq}).
Specifically, we compute the chi matrix, $\chi$, for process ${\mathcal E}(\rho)$, by employing the Choi-Jamiolkowski isomorphism~\cite{Gilchrist:2005p062310} and following the computational procedure described in Ref.~\cite{Frees:2019p73}.
Since the single-qubit gates are fast, compared to the sideband-mediated gates, we simply assume they can be performed perfectly and without noise.
As in Ref.~\cite{Frees:2019p73}, we are only concerned with initial states in the subspace $\{\ket{n_{1}n_{2}}\otimes\ket{0}_{r1}\}$, where $\{n_{1},n_{2}\}=\{0,1\}$ are the logical basis states of qubits 1 and 2, and $\ket{0}_{r1}$ is the ground state of the resonator.
Within the Choi-Jamiolkowski formalism, we then compute the process matrix $\chi=d\,\rho_{\mathcal E}$, where $d=4$ is the dimension of the subspace of interest, and $\rho_{\mathcal E}$ is the density matrix of a system comprised of the two qubits (1 and 2) and the cavity ($r1$), as well as two replicated qubits (3 and 4) coupled to a replicated cavity ($r2$)~\cite{Gilchrist:2005p062310}.
Here, the initial state for the simulations is given by $\rho_{\mathcal E}(0)=\ket{\Phi}\bra{\Phi}$, where
$\ket{\Phi}=\frac{1}{2}\sum_{n_i}^{0,1} \ket{n_{1}n_{2}}\otimes\ket{0}_{r1}\otimes\ket{n_{3}n_{4}}\otimes\ket{0}_{r2}$, and the full system evolution is governed by
\begin{multline}
\dot{\rho}_{\mathcal E}=-\frac{i}{\hbar}[I_{3,4}\otimes I_{r2}\otimes H_\text{driven},\rho_{\mathcal E}] \\
+\sum_j^{1,2,3,4}\Big[ \frac{1}{2T_2^*}(\sigma_{z}^{(j)}\rho_{\mathcal E}\sigma_{z}^{(j)}-\rho_{\mathcal E})  \\
+ \frac{1}{2T_1}(2\sigma_{01}^{(j)}\rho_{\mathcal E}\sigma_{10}^{(j)}-\sigma_{10}^{(j)}\sigma_{01}^{(j)}\rho_{\mathcal E}-\rho_{\mathcal E}\sigma_{10}^{(j)}\sigma_{01}^{(j)}) \\
+ \frac{1}{2T_{2\rho}}(\sigma_{01}^{(j)}\rho_{\mathcal E}\sigma_{10}^{(j)}+\sigma_{10}^{(j)}\rho_{\mathcal E}\sigma_{01}^{(j)}-\rho_{\mathcal E})  \\
+ \frac{\kappa}{2}(2a\rho_{\mathcal E} a^\dagger-a^\dagger a\rho_{\mathcal E}-\rho_{\mathcal E} a^\dagger a) \Big] , \label{eq:Master2}
\end{multline}
in analogy with Eq.~(\ref{eq:Master}), where $I_{3,4}$ and $I_{r2}$ represent identity matrices for the replicated qubits and resonator.
The process fidelity is finally computed using $F=\text{Tr}(\chi_\text{ideal}\chi)$, where $\chi_\text{ideal}$ is the ideal process chi matrix for a CZ-gate operation, defined in the absence of noise. 

We wish to compute CZ-gate fidelities as a function of various system parameters.
However, there are many parameters that may be considered, including $\varepsilon$, $\tau_1$, $\tau_2$, $E_\text{ST}$, $g_0$, $\omega$, $\omega_r$, $A_\varepsilon$, $A_\tau$, and $A_{E_\text{ST}}$.
We therefore begin by setting two of the parameters, $\overline E_\text{ST}/h=$10~GHz~\citep{Thorgrimsson:2017p32} and $g_0/2\pi=60$~MHz~\citep{borjans2020split}, to typical experimental values, since these properties are not broadly tunable after fabrication is complete, although some fine tuning is possible. 
We additionally satisfy the following constraints, which were described in previous sections.
First, we ensure that $\varepsilon$ is tuned to a sweet spot, as discussed in Sec.~\ref{sec:optimal}, with  $\varepsilon=\varepsilon_\text{CSS}(\tau_1,\tau_2)$ for the CSS sweet spot, or $\varepsilon=\varepsilon_\text{VSS}(\tau_1,\tau_2)$ for the VSS sweet spot.
Next, for transverse sideband operation, we satisfy the constraints $\Omega_R^{(j)}=\Delta^{(j)}$.
For longitudinal sidebands, we satisfy $\omega^{(j)}=\Delta^{(j)}$ and set $\delta\omega/\omega^{(j)}\approx 1.84$, where $\hbar\delta\omega\approx A_\text{ST,max} \leq 4$~GHz, as discussed in Sec.~\ref{sec:2QH}.
(For simplicity, here, we assume $\Delta^{(1)}=\Delta^{(2)}$.)
For $\tau$-driving, we choose $A_{\tau,\text{max}}=\sqrt{\delta\tau_1^2+\delta\tau_2^2}=\sqrt{\tau_1^2+\tau_2^2}/5$, as mentioned above.
For all driving schemes, we limit our search to the range $\tau_1/h,\tau_2/h,\omega_r,A_{\varepsilon,\text{max}}\leq 10$~GHz, which is typical for QDHQs~\citep{Thorgrimsson:2017p32}.
After applying these constraints, we are left with a more manageable set of parameter sweeps.
However, $\varepsilon$-driving still involves more free parameters than $\tau$-driving or $E_\text{ST}$-driving; we therefore limit our investigation to several discrete values of $A_\varepsilon$.
The CZ-gate fidelity estimates described in Sec.~\ref{sec:results}, below, are reported after finally optimizing over a full sweep of the remaining, unconstrained tuning parameters.

\begin{figure*}[t]
\includegraphics[width=1\textwidth]{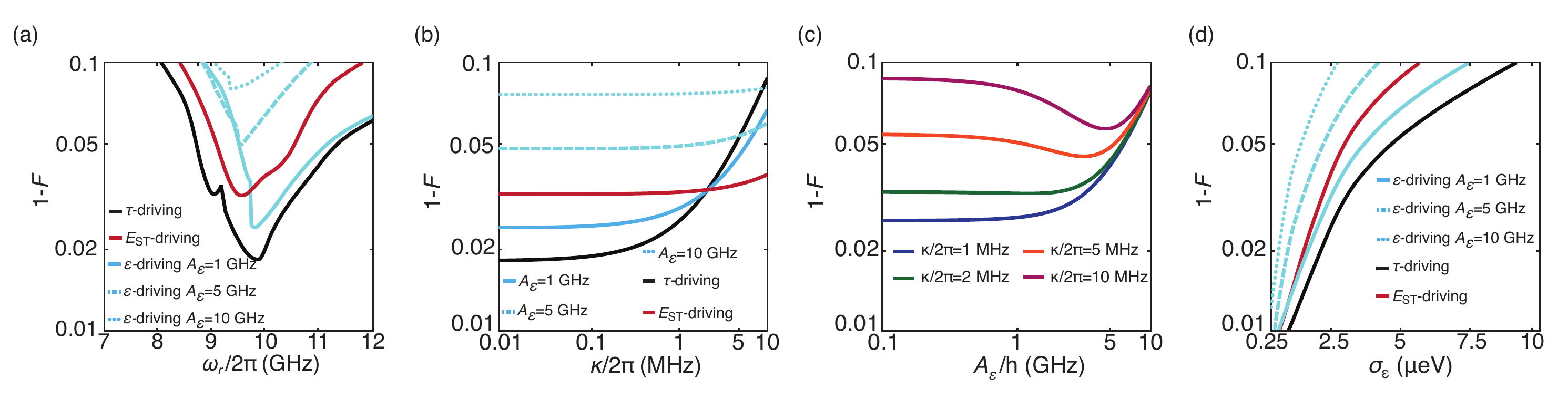}
\caption{
Infidelities ($1-F$) of two-qubit
CZ gates,
%Two-qubit CZ-gate infidelities,
optimized over a wide range of tuning parameters, in
the presence of $1/f$ charge noise with amplitude
$\sigma_\varepsilon=2$~$\mu$eV.
(a) Infidelity as a function of the resonator frequency, $\omega_r$, assuming the cavity decay rate $\kappa=0$.
Results are shown for $\tau$, $E_\text{ST}$, and $\varepsilon$-driving.
Sudden changes in the slope occur when the optimal fidelity switches from a CSS (small $\omega_r$) to a VSS (large $\omega_r$).
(b) Infidelity as a function of $\kappa$,
%the cavity decay rate, $\kappa>0$,
optimizing over all variables, including $\omega_r$.
(c) Infidelity as a function of $A_\varepsilon$, for $\varepsilon$-driving and several values of $\kappa$. 
(d) Infidelity as a function of $\sigma_\varepsilon$, the standard deviation of the charge noise, with $\kappa=0$.
}
\label{fig4}
\end{figure*}

\section{Results} \label{sec:results}
The main results of our two-qubit gate simulations are summarized in Fig.~4. 
We begin by calculating the maximum fidelities, $F$ (or minimum infidelities, $1-F$), for the CZ-gating protocol described in Sec.~\ref{sec:transverse}, by applying the $\varepsilon$, $\tau$, or $E_\text{ST}$-driving schemes.
To first characterize the qubit, we include only $T_2^*$, $T_1$, and $T_{2\rho}$ decoherence processes in our simulations, setting the cavity decay rate to $\kappa=0$.
Specifically, we use $T_2^*$ and $T_{2\rho}$ values obtained from our charge-noise simulations in Secs.~\ref{sec:CNsims} an \ref{sec:Rabiflucts}, assuming a charge noise level of $\sigma_\varepsilon=2$~$\mu$eV.
These results may be considered as upper bounds on the fidelity, consistent with very high-$Q$ (low-$\kappa$) cavities, which could become available in the future.
The resulting infidelities are plotted in Fig.~4(a) as a function the cavity frequency, $\omega_r$.
Note that $\omega_r$ affects the two-qubit evolution directly, as indicated in Eq.~(\ref{eq:Hdriv}), but also indirectly through the sideband constraints.

For $\kappa=0$, we thus find that the highest fidelities occur for $\tau$-driving, with a maximum fidelity of $F\approx 98.1$\%.
To understand why $\tau$-driving is preferred, we first note that transverse gates (obtained by $\varepsilon$-driving or $\tau$-driving) are generally faster than longitudinal gates (obtained by $E_\text{ST}$-driving), because they occur closer to the charging transition ($\varepsilon\approx 0$), where the dipole matrix element is large; in turn, this allows for shorter exposure times to charge fluctuations.
Secondly, we note that the tunnel-coupling noise strength, $\sigma_\tau$, is much smaller than the detuning noise strength ($\sigma_\tau\approx 10^{-3} \sigma_\varepsilon$~\citep{huang2018spin}), yielding negligible Rabi noise for $\tau$-driving compared to $\varepsilon$-driving.
Based on the trends observed in Fig.~4(a), it appears that better fidelities could potentially be obtained for $\varepsilon$-driving, using smaller $A_\varepsilon$ values than the ones considered here; this is because the coherence-limiting mechanism in Eq.~(\ref{eq:T2rho}) scales as $T_{2\rho}\propto A_\varepsilon^{-1}$. 
However, a smaller $A_\varepsilon$ with $\kappa>0$ (studied below) suffers from cavity decay, and does not provide better fidelities.

For $\kappa=0$, the optimal cavity frequencies occur at the locations where the infidelities in Fig.~4(a) are minimized.
For each of the driving schemes, the optimal values of $\omega_r$ are found to occur near 10~GHz, which can be understood from the following arguments.
Since the best sweet spots are obtained in the far-detuned regime, where $\hbar\omega_q\approx E_\text{ST}$, the optimal cavity frequency must therefore satisfy $\omega_r/2\pi\approx\omega_q/2\pi\approx E_\text{ST}/h\approx 10$~GHz.
Moving away from this optimal value requires using sub-optimal sweet spots, which in turn reduces the fidelity.
For $\tau$-driving, it is difficult to operate QDHQs outside the range $A_{\tau,\text{max}}=\sqrt{\delta\tau_1^2+\delta\tau_2^2}\leq\sqrt{\tau_1^2+\tau_2^2}/5$ or $\tau_1/h,\tau_2/h\leq 10$~GHz, yielding driving amplitudes that are relatively weak.
Moreover, the dipole matrix element for $\tau$-driving is weak in the far-detuned regime, and the resulting $\Omega_R$ is small.
Satisfying the side-band constraint, $\Delta^{(j)}=\Omega_R^{(j)}$, we again find a small qubit-cavity detuning, $\Delta^{(j)}$, such that $\omega_r/2\pi\approx 10$~GHz for optimized operations.
In contrast, the restriction to small driving amplitudes does not apply to $\varepsilon$-driving, since $A_\varepsilon$ can be quite large.
This suggests optimal $\omega_r/2\pi$ values that are below 10~GHz.
However, such strong driving also increases the decoherence arising from the $T_{2\rho}$ mechanism.
Hence, optimal $\omega_r/2\pi$ values for $\varepsilon$-driving (for $\kappa=0$) are also near 10~GHz.
In principle, we could increase $T_{2\rho}\propto A_\varepsilon^{-1}$ by employing driving amplitudes $A_\varepsilon$ that are smaller than those considered in Fig.~4(a).
However, this would require cavity frequencies $\omega_r/2\pi >10$~GHz, which are difficult to achieve in the laboratory.
Finally, $E_\text{ST}$-driving allows for optimal resonator frequencies that are somewhat further away from $\omega_r=10$~GHz.
This is because relatively large driving amplitudes are possible, $\hbar\delta\omega\approx A_{E_\text{ST},\text{max}}\leq 4$~GHz, as described in Appendix~B.
The coupling requirement $\delta\omega/\omega^{(j)}\approx 1.84$ then suggests that the resulting qubit driving frequencies $\omega^{(j)}$ can be of order GHz.
Satisfying the sideband constraint $\Delta^{(j)}=\omega^{(j)}$ therefore allows for large qubit-cavity detunings, and optimized operation with smaller $\omega_r$ values.

As indicated in Figs.~2(c) and 2(d), the best $\tau_1$ and $\tau_2$ values for two-qubit gates are located near the boundary between the regions with zero and two sweet spots, shown in Fig.~2(a).
As explained in Sec.~\ref{sec:optimal}, this boundary represents a line of second-order sweet spots, where dephasing effects are suppressed.
Unfortunately, leakage effects are also enhanced in this region.
Consequently, optimal fidelities are obtained slightly away from the boundary line.
The best overall fidelity for $\kappa=0$ is given by $F=98.1\%$, and occurs at the $\tau_1,\tau_2$ values marked by a purple star in Fig.~2(a), which is almost a second-order sweet spot. 
At this setting, a transversally-driven red-sideband CZ gate is implemented in about 30~ns.
Although two sweet spots (VSS and CSS) are observed throughout the green region in Fig.~2(a), we find that better fidelities are obtained from a VSS, rather than a CSS, since it occurs further from the charging transition and therefore enjoys better coherence properties. 
Consequently, the optimal behavior at the purple star corresponds to a VSS.
In contrast, the orange region in Fig.~2(a) has only one sweet spot, which is of the CSS type.
Therefore, upon crossing from the green region to the orange region, the optimal behavior switches from a VSS to a CSS, as revealed by a sudden jump in the infidelity slope in Fig.~4(a).

We now include the effects of cavity decay by setting $\kappa>0$. 
For each value of $\kappa$, we compute the infidelity as a function of $\omega_r$, as in Fig.~4(a), and identify the optimal values.
These optimized fidelities are plotted in Fig.~4(b) as a function of typical $\kappa/2\pi$ values in the range 0.01-10~MHz, where $\kappa/2\pi =10$~MHz corresponds to a resonator quality factor of $Q\approx 10^3$.
In the limit of $\kappa\approx 0$, $\tau$-driving is found to give the best fidelity, as consistent with Fig.~4(a).
Larger $\kappa$ values reduce the fidelity for all the driving schemes, due to the Purcell effect; however, the size of the effect depends on the driving scheme.
Most importantly, the Purcell effect is enhanced when the qubit and cavity are near resonance.
Since optimized $\tau$-driving must be performed near resonance, as discussed above, this scheme suffers the most greatly when $\kappa$ is large.
Its fidelity can be improved slightly by increasing $A_\tau$, which increases $\Delta$ and suppresses the Purcell effect.
However, we have also noted that $A_\tau$ cannot be increased significantly.
We find that $\varepsilon$-driving outperforms $\tau$-driving when $\kappa/2\pi >2$~MHz, because $A_\varepsilon$ can be significantly increased, to maintain a large qubit-cavity detuning.
On the other hand, a large $A_\varepsilon$ also tends to suppress the fidelity, due to the effect of Rabi-frequency fluctuations.
Hence, an optimal value of $A_\varepsilon$ emerges in the simulations, as a function of $\kappa$, as shown in Fig.~4(c).
(We note again that small values of $A_\varepsilon/h\lesssim 1$~GHz suggest using large optimized $\omega_r$ values, which are difficult to achieve in the laboratory.)
$A_{E_\text{ST}}$ can also be increased, to keep the qubit and cavity well detuned.
However, in the large-$\varepsilon$ regime, the dipole matrix element for $E_\text{ST}$-driving is much larger than for $\varepsilon$-driving, so it is not necessary to strongly drive $E_\text{ST}$.
As a result, a large qubit-cavity detuning can be maintained without inducing detrimental Rabi-frequency fluctuations.
Consequently, $E_\text{ST}$-driving is found to give the best fidelities in Fig.~4(b), when $\kappa/2\pi >2$~MHz.
 
Up to this point, we have considered a fixed level of charge noise, $\sigma_\varepsilon=2$~$\mu$eV.
Reducing the noise naturally leads to improvements in the gate fidelity.
To characterize the magnitude of such improvements, we close this section by repeating the analyses described above for a range of $\sigma_\varepsilon$ values.
To focus on the charge noise rather than cavity decay, we assume a perfect resonator, with no Purcell decay ($\kappa=0$), obtaining the results shown in Fig.~4(d).
As expected, the infidelity grows rapidly with $\sigma_\varepsilon$. 
The value of $\sigma_\varepsilon$, below which the fidelity exceeds 99$\%$, corresponds to 0.5~$\mu$eV, 0.6~$\mu$eV, or 0.9~$\mu$eV, for $\varepsilon$-driving, $E_\text{ST}$-driving, or $\tau$-driving, respectively.
We note that a noise level of $\sigma_\varepsilon<0.9$~$\mu$eV has previously been demonstrated in Si qubits~\cite{Mi:2018p161404}.

\section{Conclusions}
In this work, we have focused on the physics of two-qubit gates between QDHQs.
We now conclude by describing more general operating modes.
In particular, we explain how to turn the qubit-cavity coupling off, and how to tune the qubit into a regime appropriate for single-qubit gates.

An important benefit of sideband gating is that it can largely be extinguished by turning off the drive.
However, a residual dispersive coupling remains, resulting in unwanted dynamics that evolve at a rate proportional to $g^2/\Delta$.
This coupling can then be suppressed by adiabatically tuning $g(\varepsilon)=g_0\, d_{01}(\varepsilon)$ to zero, making use of the fact that the dipole moment $d_{01}$ is strongly suppressed when $\varepsilon\gtrsim \tau_0,\tau_1,E_\text{ST}$.

A bigger challenge for controlling the coupling, and for performing other types of QDHQ gates, is to be able to tune $\varepsilon$ while remaining protected from charge noise at a sweet spot.
The sweet spots depicted in Fig.~2(a) satisfy the nontrivial relations $\varepsilon=\varepsilon_\text{CSS}(\tau_1,\tau_2)$ or $\varepsilon=\varepsilon_\text{VSS}(\tau_1,\tau_2)$.
It is usually difficult to vary $\tau_1$ and $\tau_2$ independently, because their ratio is largely determined by atomistic details of the quantum well interface, while their average value is controlled by a single tunnel-barrier top gate.
Limited, independent control of $\tau_1$ and $\tau_2$ has previously been achieved using plunger-gate controls to modify the shape and/or the position of the dot in the quantum well~\cite{Abadillo-Uriel:2018p165438}.
An easier and more conventional approach is to leave the ratio $\tau_1/\tau_2$ fixed~\cite{Yang:2020,Borjans:2021preprint}, varying only the tunnel-barrier height.
The resulting tuning paths correspond to straight lines in Fig.~2(a); the requirement for ensuring a sweet spot is then to properly adjust $\varepsilon$.

To navigate Fig.~2(a) in this way, when performing gate operations, we suggest a strategy similar to the one proposed for singlet-triplet qubits in Ref.~\cite{abadillo2019enhancing}.
In this approach, the qubit is initially tuned to a desirable starting point on the line defined by $\tau_1/\tau_2=\text{const.}$
Let us assume that $\tau_1>\tau_2$, so that the line traverses the green region of Fig.~2(a), which contains two sweet spots.
Gating then proceeds by simultaneously adjusting the tunnel-barrier gate voltage, which controls the barrier height, and the plunger-gate voltage, which controls the detuning, to remain at a sweet spot.
This allows us to maneuver from a location in the interior of the green region to the boundary of the blue region, which corresponds to a second-order sweet spot.
The latter is an excellent location for idling. 
The second-order sweet spot also represents the merging of the CSS and the VSS.
From this branch point, we may therefore follow either a CSS or a VSS path, which allows us to make the best use of the different sweet spots -- the CSS being advantageous for faster gates and the VSS being advantageous for quieter gates. 
The most desirable line of sweet spots passes through the optimal working point, marked by a purple star in Fig.~2.
Unfortunately, the tunnel coupling ratio $\tau_1/\tau_2$ is the most difficult system parameter to control, and it may not always be possible to achieve this target.
We have therefore performed additional simulations to study such control errors.
Assuming errors in the range of 5-25$\%$ for $\tau_1/\tau_2$, and additional errors of 1$\%$ for $\varepsilon$, we observe a 1-15$\%$ reduction of the fidelity, when aiming to operate at the ideal working point. 

In summary, we have investigated the fidelity of two-qubit gates in QDHQs, mediated by a microwave cavity.
To mitigate the Purcell effect in this system, it is necessary to detune the qubit and cavity into the dispersive regime.
We focus on sideband gating protocols in which one of the QDHQ control parameters, $\varepsilon$, $\tau$, or $E_\text{ST}$, is driven, with stronger drives enabling greater dispersiveness.
Additional benefits of sideband gating include (1) potentially high gate speeds, (2) the resonance condition can be achieved by tuning the driving frequency, which is much easier than tuning the qubit frequency, (3) the coupling can nominally be extinguished by simply turning off the drive.
Optimal performance is achieved by working at either a VSS or a CSS, whose locations we have mapped out.
In the regime where both types of sweet spots are present, we achieve higher fidelities at a VSS.
For most cases of interest, the desirable sweet spots are located in the far-detuned regime, where $\varepsilon\gtrsim \tau_1,\tau_2, E_\text{ST}$, and the qubit energy is given by $\hbar\omega_q\approx E_\text{ST}$.

Several competing effects suppress the two-qubit gate fidelities.
In the strong-driving regime, Rabi-frequency noise ($T_{2\rho}$) is important, although it is typically only relevant for $\varepsilon$-driving. 
The Purcell effect is also enhanced for weak driving because the resulting qubit-cavity detuning is small.
Consequently, the Purcell effect is important for $\tau$-driving, because these parameters typically cannot be strongly driven. 
To balance these competing noise effects, we optimize the two-qubit gate fidelity over many different tuning parameters.
For typical charge noise levels of $\sigma_\varepsilon\approx 2$~$\mu$eV, and cavity decay rates of $\kappa/2\pi\approx 2.1$ MHz, we obtain the best results for $E_\text{ST}$-driving, with gate fidelities of 96.6$\%$.
For two-qubit gates, the Purcell effect is currently the most serious obstacle for high-fidelity operation.
Technical improvements in resonator quality factors would therefore be welcome for this system.
To estimate the possible improvements, we consider the limiting case, $\kappa=0$, for which the best results are obtained for $\tau$-driving, yielding gate fidelities of about 99\% for low charge noise $\sigma_\varepsilon<1$~$\mu$eV, which has already been achieved in the laboratory.

\begin{acknowledgements}
We are grateful to Mark Eriksson for many illuminating discussions.
This work was supported in part by the Army Research Office (W911NF-17-1-0274) and by the Vannevar Bush Faculty Fellowship program sponsored by the Basic Research Office of the Assistant Secretary of Defense for Research and Engineering and funded by the Office of Naval Research through Grant No.\ N00014-15-1-0029. The views and conclusions contained in this document are those of the authors and should not be interpreted as representing the official policies, either expressed or implied, of the U.S. Government. The U.S. Government is authorized to reproduce and distribute reprints for Government purposes notwithstanding any copyright notation herein.
\end{acknowledgements}

\appendix

\section{Noise-Induced Leakage} \label{sec:leakage}
\begin{figure*}[t]
\includegraphics[width=1\textwidth]{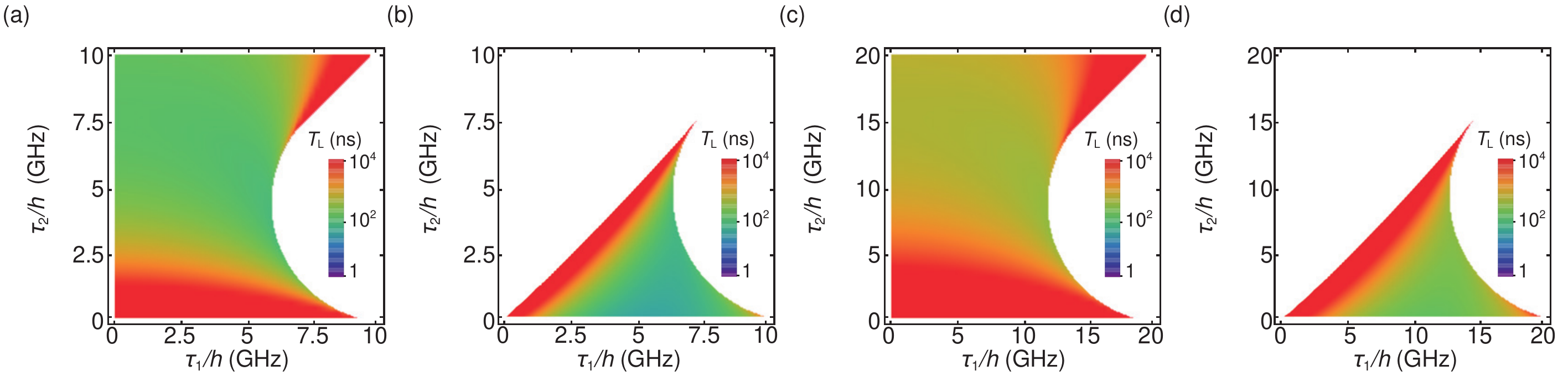}
\caption{Simulations of noise-induced leakage rates, $1/T_L$, as described in Appendix~A. 
In all cases, we find that noise-induced leakage is dominated by other noise or leakage effects, suggesting that it is not necessary to include noise-induced leakage in our two-qubit gate simulations.
(a) Results for $E_\text{ST}/h=10$~GHz at a CSS.
(b) Results for $E_\text{ST}/h=10$~GHz at a VSS.
(c) Results for $E_\text{ST}/h=20$~GHz at a CSS.
(d) Results for $E_\text{ST}/h=20$~GHz at a VSS.}
\label{fig5}
\end{figure*}

Our two-qubit gate simulations of QDHQ include the effects of coherent leakage to state $\ket{L}$, induced by driving.
The presence of charge noise, Eq.~(\ref{eq:noise}), can also cause undesired transitions to the leakage state. 
To estimate the importance of this effect, we perform charge-noise simulations as described in Sec.~\ref{sec:CNsims}, without any additional driving term.
For the initial state, we choose the fully relaxed state $\rho=1/2(\ket{0}\bra{0}+\ket{1}\bra{1})$.
We then monitor the time dependence of the ensemble-averaged density matrix, $\overline \rho$.
The simulation includes all three quantum-dot states, $\ket{0}$, $\ket{1}$, and $\ket{L}$, and we estimate the leakage rate, $1/T_L$, by fitting the results to 
\begin{equation}
\overline\rho_{00}(t)+\overline\rho_{11}(t)=\frac{2+\exp(-t/T_L)}{3}.
\end{equation}

The results of such leakage simulations are shown in Figs.~5(a) and 5(b) for $E_\text{ST}=10$~GHz.
Generally, we find that $T_L\gg T_2^*$, suggesting that there is no reason to include noise-induced leakage in our two-qubit simulations.
The only regime where the ratio $T_L/T_2^*$ drops below 10 is very near the line of second-order sweet spots.
However, along this line, all leakage effects are enhanced, including driving-induced leakage, which dominates over the noise-induced leakage.

Finally, to test whether leakage can be suppressed for $E_\text{ST}>10$~GHz, we repeat the leakage simulations for the case $E_\text{ST}/h=20$~GHz.
The results shown in Figs.~5(c) and 5(d) confirm than any unwanted noise-induced leakage is strongly suppressed by increasing $E_\text{ST}$.

\section{Estimation of the $E_\text{ST}$-driving amplitude, $A_{E_\text{ST}}$} 
\label{sec:vsdriving}
The valley splitting in SiGe/Si/SiGe quantum wells is determined by the details of the penetration of the electron wavefunction into the SiGe barriers~\citep{Friesen:2007p115318}. 
The splitting is approximately proportional to the electric field perpendicular to the quantum well.
The plunger gate voltage therefore provides an effective knob for tuning the valley splitting; however, it also controls the dot occupation, effectively limiting the range of valley splittings that can be obtained in a qubit.
In a recent valley-splitting experiment, it was shown that a typical range of electric fields for a SiGe/Si/SiGe quantum dot in which the dot preserves the same configuration is given by  $F_z\approx (2.8\pm 0.1)$~MV/m~\citep{Tom_unpublished}.

To estimate $A_{E_\text{ST}}$, we perform a tight-binding simulation of a quantum dot in a SiGe/Si/SiGe quantum well, using the methods described in Refs.~\cite{Abadillo-Uriel:2018p165438} and \cite{Boykin:2004p165325}.
This method also allows us to account for the atomic-scale step disorder at the quantum well interface, which plays an important role in determining the magnitude of the valley splitting.
In our simulations, we consider the electric field range given above, and adjust the disorder step width to match the average valley splitting used in this work, $E_\text{ST}=10$~GHz.
The result of this procedure gives $A_{E_\text{ST}}=4$~GHz.
We note that, for any driving scheme, including $E_\text{ST}$-driving, varying a single gate voltage (in this case, the plunger gate) can affect multiple system parameters, including the valley splitting, the detuning, and the tunnel coupling. 
However, it is common practice nowadays to orthogonalize the control of these parameters by adjusting several gate voltages simultaneously, using a procedure known as ``compensation."

\bibliography{hybrid}
\end{document}